\documentclass{article}
\usepackage{waspaa21,amsmath,graphicx,url,times}
\usepackage{amssymb}
\usepackage{amsfonts}
\usepackage{mathtools}
\usepackage{multirow}
\usepackage{bm}
\usepackage{nicefrac}
\usepackage{tabularx}
\usepackage{booktabs}
\usepackage{xcolor}
\usepackage{pifont}
\usepackage{amsthm}
\usepackage{extarrows}
\usepackage{hyperref}
\usepackage{balance}
\usepackage{subcaption}
\usepackage{caption}
\usepackage[ruled,vlined]{algorithm2e}

\usepackage{array}

\SetKwInput{KwInput}{Input}                
\SetKwInput{KwOutput}{Output}              

\def\feddataset{\textit{LibriFSD50K}}
\def\feddatasetSpace{\textit{LibriFSD50K} }
\def\fedenhance{\textsc{FedEnhance}}
\def\fedenhanceSpace{\textsc{FedEnhance} }

\def\w{{\bm{\theta}}}
\def\y{{\mathbf y}}
\def\s{{\mathbf s}}

\def\L{{\cal L}}
\def\D{{\cal D}}
\def\R{{\mathbb R}}

\def\R{\mathbb{R}}

\def\s{\mathbf{s}}
\def\es{\widehat{\mathbf{s}}}

\makeatletter
\def\hlinewd#1{%
  \noalign{\ifnum0=`}\fi\hrule \@height #1 \futurelet
   \reserved@a\@xhline}
\makeatother

\title{Separate but Together: Unsupervised Federated Learning \\for Speech Enhancement from non-IID Data}

\name{Efthymios Tzinis$^{1}$\thanks{Code: \texttt{https://github.com/etzinis/fedenhance}
}, \quad
      Jonah Casebeer$^{1}$, \quad
      Zhepei Wang$^{1}$, \quad
      Paris Smaragdis$^{1, 2}$}
\address{$^1$ University of Illinois at Urbana-Champaign, Department of Computer Science, Urbana, IL, USA\\
$^2$ Adobe Research, San Jose, CA, USA\\
}

\begin{document}

\ninept
\maketitle

\begin{sloppy}

\begin{abstract}
We propose \fedenhance, an unsupervised federated learning (FL) approach for speech enhancement and separation with non-IID distributed data across multiple clients. We simulate a real-world scenario where each client only has access to a few noisy recordings from a limited and disjoint number of speakers (hence non-IID). Each client trains their model in isolation using mixture invariant training while periodically providing updates to a central server. Our experiments show that our approach achieves competitive enhancement performance compared to IID training on a single device and that we can further facilitate the convergence speed and the overall performance using transfer learning on the server-side. Moreover, we show that we can effectively combine updates from clients trained locally with  supervised and unsupervised losses. We also release a new dataset \feddatasetSpace and its creation recipe in order to facilitate FL research for source separation problems. 
\end{abstract}


\section{Introduction}
\label{sec:intro}
Recent advances in deep learning have enabled the development of neural network architectures capable of separating individual sound sources from mixtures of sounds with high fidelity. Discriminative separation models with supervised training have obtained state-of-the-art performance on multiple tasks such as music separation \cite{takahashi2020d3netMusicSeparationSOTA}, speech separation \cite{nachmani2020voiceUknownNumberOfSpeakers, subakan2020attentionSeparation} and speech enhancement \cite{wang2020complexSpectralMappingSpeechEnhancementRobustASR, Isik2020PoCONetSpeechEnhacement}. However, gathering clean source waveforms to perform supervised training under various domains can be cumbersome or even impossible. Other works have focused on less supervised approaches by leveraging contrastive learning \cite{sivaraman2021personalizedSpeechEnhancement} or using weak labels containing sound-class information \cite{pishdadian2020findingSeparationWeakLabels, kong2021speechEnhancementWeakLabels}. Moreover, unsupervised approaches using spatial information from multiple microphones \cite{tzinis2019unsupervised,seetharaman2019bootstrapping,drude2019unsupervised} and visual cues \cite{gao2019co} have also shown promising results. Mixture invariant training (MixIT) has shown great potential for single-channel sound separation and speech enhancement by training on synthetic mixtures of mixtures \cite{wisdom2020MixIT}. Despite that these works lessen the reliance on supervised data, they still require huge consolidated audio data collections being available for IID training on a single device.


Federated learning (FL) \cite{FLkonevcny2016StrategiesForImprovingCommunication} has provided a distributed and privacy preserving framework where each client trains on their local data and communicates updates to a central server. The central server aggregates those updates and distributes a new model at each communication round. Several studies have shown that by repeating this process, one can aptly train a global model without violating client privacy \cite{FLmcmahan2017firstFedAvg} even under cases where data are not distributed IID to the clients \cite{FLsattler2019FLfromNonIID}. FL has also been applied to train audio models for keyword spotting \cite{FLleroy2019KeywordSpotting}, automatic speech recognition \cite{FLguliani2020trainingASRwithFL} and sound event detection \cite{FLjohnson2021SoundEventDetection}. Nevertheless, the aforementioned approaches, as well as most FL setups, require supervised data to be available on the client side and are even less successful when multiple clients are present or the IID assumption is violated \cite{FLkairouz2019advancesAndOpenProblems}. The aforementioned problem led to the development of recent FL algorithms which require less supervision \cite{van2020towardsunsupFL, servetnyk2020unsupervisedFL}.

In this work, we tackle the real world problem of learning a speech enhancement model where a central server (e.g. a company) aggregates updates across numerous clients. The local client datasets contain noisy mixtures and may greatly vary across clients. Thus, we present \fedenhance{}, an FL system which is capable of learning a separation model for speech enhancement without relying on several common assumptions such as: a) requiring supervised data and b) assuming IID distribution of the data across the clients. In essence, each client has access to a limited number of noisy speech mixtures and isolated noise recordings. All clients perform local unsupervised training using MixIT by synthesizing a mixture from a noisy speech recording and a noise recording. To evaluate our approach, we introduce a realistic and challenging speech enhancement dataset, namely, \feddatasetSpace containing around $50$ hours of training data and with $280$ speaker IDs. We use the speaker variability for simulating real-world situations that clients only have access to noisy speech recordings from one speaker (e.g. a smartphone client can easily record themselves). We analyze the convergence behavior of \fedenhanceSpace{} and show that our approach can scale to multiple clients. Moreover, we show that we can expedite the convergence and boost the overall performance of our FL method by transferring knowledge from another medium-size speech enhancement dataset. Finally, our experimental results show that we can effectively combine updates from clients with supervised and unsupervised data using different loss functions.  

\section{Unsupervised federated learning for speech enhancement}
\label{sec:method}

\fedenhanceSpace{} follows a conventional FL setup with one central server orchestrating the overall communication and $C$ clients which perform local training on their private data $\{\D_c\}_{c = 1, \dots, C}$. 

\subsection{Server-clients communication}
\label{sec:method:communication}
The server owns a global copy of the model weights $\w^{(r)}_g$ which are distributed to the decentralized clients at each round $r$. Formally, we assume that the server shares the same separation network architecture with the clients  $f(\cdot; \w^{(r)}_{g})$ which is parameterized through the global weights $\w^{(r)}_{g}$ at the communication round $r$. For an input mixture with $T$ samples in the time-domain $\mathbf{x} \in \R^{T}$, each separation model outputs $M$ sources, namely, $\es = f(\mathbf{x}; \w) \in \R^{M \times T}$. At the update step, the $c$-th client from the set of available clients $\mathcal{A}^{(r)}$ provides the updated weights $\w_c^{(r)}$ after training independently on its private data $\mathcal{D}_c$. Finally, the server aggregates those weights in order produce the updated model $\w^{(r+1)}_g$. The new model is going to be distributed again across all clients and the process is repeated for $R$ communication rounds as described in Algorithm \ref{alg:fedavg}.

\begin{algorithm}[htb!]
\SetAlgoLined
\KwInput{$\w^{(0)}_{g}$ \tcp{Initial server model weights}}
\KwOutput{$\w^{(R)}_{g}$} 
 \For{$r = 1$; $r{+}{+}$; while $r <= R$}{
 \tcp{Available clients $\mathcal{A}^{(r)} \subseteq \{1, \dots, C\}$}
 \For{$n \in \mathcal{A}^{(r)}$}{
 \tcp{Server distributes the model}
 $\w^{(r)}_c \gets \w^{(r-1)}_g$ \\
 \tcp{Local training on private $\D_c$}
 $\w^{(r)}_c \gets \textsc{ClientUpdate}(\w^{(r)}_c, \D_c)$
 }
 \tcp{Server update for round $r$}
 $\w^{(r)}_g \gets \frac{1}{|\mathcal{A}^{(r)}|} \sum_{n \in \mathcal{A}^{(r)}} \w^{(r)}_n$}
 \caption{\textsc{FedAvg} \cite{FLmcmahan2017firstFedAvg} with a server and $C$ clients.}
 \label{alg:fedavg}
\end{algorithm}

\subsection{Local training on the client-side}
\label{sec:method:local_training}
We present how each client performs local training on its private data for the task of speech enhancement. We assume that the $c$-th client has access only to its private data $\D_c$ which consist of two portions $\D_c = (\D_c^{m}, \D_c^{n})$. Specifically, $\D_c^{m}$ is the part of the data that contains mixtures of speech and noise while $\D_c^{n}$ contains only clean noise. Following the training procedure presented in \cite{wisdom2020MixIT}, each client generates artificial mixtures of mixtures (MoMs) $\mathbf{x} = \mathbf{s} + \mathbf{n}_1 + \mathbf{n}_2$ using a noisy speech example $\mathbf{m}=\mathbf{s} + \mathbf{n}_1 \sim \D_c^{m}$ and a clean noise recording $\mathbf{n}_2 \sim \D_c^{n}$. The separation model always estimates $M=3$ sources. We distinguish two different cases for each client's private data $\D_c^{m}$ while we always assume that each client has access to clean noise recordings $\mathbf{n}_2 \sim \D_c^{n}$.

\textbf{Supervised data}:
A supervised client would have to noisy speech files $\mathbf{m} = \mathbf{s} + \mathbf{n}_1$ as well as the isolated speech $\s$ and the noise $\mathbf{n}_1$ waveforms, where $(\mathbf{m}, \s, \mathbf{n}_1) \sim \D_c^{m}$ (e.g. research institutes with many hours of clean speech recordings might fall in this category). The loss function which is minimized is shown next:
\begin{equation}
\label{eq:supervised_loss}
    \begin{aligned}
    \mathcal{L}_{\operatorname{sup}} = \L(\es_1, \s) + \frac{1}{2} \underset{\pi \in \Pi_{2,3}}{\min} \left[ \L(\es_{\pi_1}, \mathbf{n}_1) + \L(\es_{\pi_2}, \mathbf{n}_2)  \right],
    \end{aligned}
\end{equation}
where $\L$ could be any signal-level loss which measures the reconstruction fidelity of each separated signal $\es_i$ w.r.t. the targets $\s$, $\mathbf{n}_1$, and $\mathbf{n}_2$. $\Pi_{2,3} = \{(2,3), (3,2)\}$ symbolizes the set of permutations used for the latter two slots $\es_2$ and $\es_3$. Thus, we train with permutation invariance w.r.t. the noise sources $\mathbf{n}_1$ and $\mathbf{n}_2$ while also forcing the model to produce the reconstructed speaker in the first slot $\es_1$.

\textbf{Unsupervised data}: This client has access only to mixtures of noisy speech $(\mathbf{m}, -) \sim \D_c^{m}$. This is the most interesting case since most of the clients would have a small data collection with noisy recordings that do not want to share directly but are willing to contribute to the FL framework by sending updated weights to the server. Now the unsupervised loss follows the mixture invariant training setup \cite{wisdom2020MixIT} and can be described as shown next:
\begin{equation}
\label{eq:unsupervised_loss}
    \begin{aligned}
    \mathcal{L}_{\operatorname{unsup}} = \underset{\pi \in \Pi_{2,3}}{\min} \left[ \L(\es_{1} + \es_{\pi_1}, \mathbf{m}) + \L(\es_{\pi_2}, \mathbf{n}_2)  \right].
    \end{aligned}
\end{equation}
We still force the model to produce the reconstructed speaker waveform at the first slot and assume that the sources in the input MoM $\mathbf{x} = \s + \mathbf{n}_1 + \mathbf{n}_2$ are independent. The assumption about clients (even unsupervised ones) having access to noise recordings $\mathbf{n}_2 \sim \D^{n}_c$ is fairly reasonable since distributing such data from the server and using them locally does not raise any privacy issues. Moreover, noise data can also be gathered independently on the client-side by running a simple voice activity detection mechanism even real-time on a modern mobile device \cite{sehgal2018realTimeVADSmartphone}.

Finally, the $c$-th client belonging to the set of the available nodes at the $r$-th communication round $c \in \mathcal{A}^{r}$ performs $K_c$ local mini-batch updates to its local weights $\w^{(r)}_c$ by minimizing one of the aforementioned loss functions in Equations \ref{eq:supervised_loss}, \ref{eq:unsupervised_loss}.

\section{Experimental Framework}
\label{sec:exp_frame}

\subsection{\feddatasetSpace Dataset}
\label{sec:exp_frame:Dataset}
To construct a realistic FL scenario, we combine speech data from the LibriSpeech dataset \cite{panayotov2015librispeech}, and noise data from FSD50k \cite{fonseca2020fsd50k}. We first split the combined 100h and 360h LibriSpeech training set into $280$ folds where each fold contains data from a unique speaker ID. Then, we remove all clips with speech related tags from both train and test portions of FSD50K. The remaining noise files are split evenly across the $280$ speaker-folds by matching four second noise segments to four second chunks of unique speech utterances. We leave the test sets from both datasets intact. \feddatasetSpace consists of approximately $50$, $3$ and $3$ hours of data for training, validation and test, respectively. To produce a training mixture with two noise sources $\s + \mathbf{n}_1 + \mathbf{n}_2$, we discard half of the speech recordings from a speaker and pair the singleton noise files $\mathbf{n}_2 \sim \D_c^{n}$ with another noisy speech mixture $\mathbf{m}=\s + \mathbf{n}_1 \sim \D_c^{m}$ from the same speaker. This procedure leads to the challenging but realistic distributions of input signal-to-noise ratio SNR shown in Figure \ref{fig:snr_distribution}.

\begin{figure}[!htb]
    \centering
      \includegraphics[width=\linewidth]{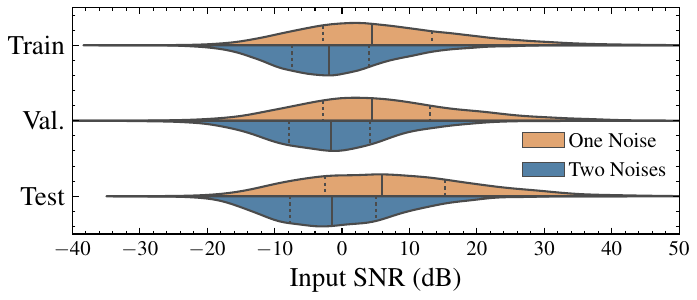}
      \caption{Input SNR distribution for the \feddatasetSpace dataset.}
      \label{fig:snr_distribution}
\end{figure}

\subsection{Separation model}
\label{sec:exp_frame:model}
Although our method can work with any separation architecture, we use a variation of the Sudo rm -rf \cite{tzinis2020sudo} separation architecture since it achieves a good trade-off between separation fidelity and time-memory computational requirements. This Sudo rm -rf variation, uses $8$ repetitive U-ConvBlocks as well as the group communication mechanism proposed in \cite{luo2020groupcomm} which divides all intermediate representations in $16$ groups of sub-bands and process them independently. By doing so, we obtain an efficient and light-weight model with only $794 \thinspace 921$ trainable parameters which can easily fit and run on an edge device with $32$ bit precision. We force the $M=3$ estimated sources to add up to the input mixture using a mixture consistency layer \cite{wisdom2019differentiableMixtureConsistency}. For all the other parameters we choose the default settings provided in \cite{tzinis2021compute} for a sampling rate of $16$k Hz.

\subsection{Training details}
\label{sec:exp_frame:train}
For our simulations, a uniformly randomly sampled set with a cardinality of $|\mathcal{A}^{(r)}|=\nicefrac{C}{4}$ clients is chosen to contribute to the update of $r$-th communication round. Each client that is available at communication round $r$ performs training on its private data $\D_c = (\D_c^{m}, \D_c^{n})$ for $K$ optimization steps equivalent to one local epoch $K_c= \lfloor \nicefrac{|\D_c|}{B} \rfloor$, where $B$ is the batch size. The signal-level loss function used (see Section \ref{sec:method:local_training}) is the negative permutation-invariant scale-invariant signal to distortion ratio (SI-SDR) \cite{le2019sdr}: 
\begin{equation}
\label{eq:SISDR}
    \begin{gathered}
    \mathcal{L}(\widehat{\y}, \y) = - \text{SI-SDR}(\widehat{\y}, \y) = - 10 \log_{10} \left( \frac{\| \alpha \y\|^2}{\| \alpha \y - \widehat{\y}\|^2} \right),
    \end{gathered}
\end{equation}
where $\alpha =  \widehat{\y}^\top  \y /\|\y\|^2$ makes the loss invariant to the scale of the estimated source $\widehat{\y} \in \R^T$ and the target signal $\y$. We train all our local-client models using the Adam optimizer \cite{adam} with an initial learning rate of $10^{-3}$ and a batch size of $B=6$ to obtain maximum parallelization on a single Nvidia GeForce RTX 2080 Ti.

\subsection{Evaluation details}
\label{sec:exp_frame:eval}
We evaluate the robustness of our learned speech-enhancement models after a communication round is complete (the server has aggregated the weights from the individual clients). We evaluate under two different noise conditions when one $\mathbf{x} = \s + \mathbf{n}_1$ or two noise sources $\mathbf{x} = \s +\mathbf{n}_1 + \mathbf{n}_2$ are active alongside the speech utterance. Specifically, we measure the SI-SDR improvement over the input mixture $\operatorname{SI-SDRi} = \operatorname{SI-SDR}(\es_1, \s) - \operatorname{SI-SDR}(\mathbf{x}, \s)$.

\subsection{Federated learning configurations}
\label{sec:exp_frame:fed_configs}
We perform several FL simulations where the speaker data are distributed in a non-IID way across the clients (which makes the setup challenging \cite{FLsattler2019FLfromNonIID} but also more realistic). Specifically, each noisy speech example $\mathbf{m} = \s + \mathbf{n}_1 \sim \D_c^{m}$ contains a unique speaker utterance and a unique noise snippet. Each client contains only utterances from certain speaker IDs which are not shared with any other client. In the case where we assume $C$ clients, we split the dataset to almost equal $\lfloor{\nicefrac{280}{C}} \rfloor$ parts according to the speaker IDs. For each local dataset, we preserve the structure of noisy mixtures $\D_c^{m}$ and noise recordings $\D_c^{n}$, as explained in Section \ref{sec:method:local_training}. We discuss in detail all our setups and how they relate to real-world scenarios.

\subsubsection{Unsupervised federated learning at scale}
\label{sec:exp_frame:fed_configs:unsup}
We want to provide a good baseline for the most challenging FL setup which is analyzed in this work, namely, developing a distributed and private system capable of learning to perform high fidelity speech enhancement where $C \in \{16, 64, 256\}$ clients learn directly from their limited noisy datasets. We analyze the convergence of the global model learned in a federated setup and we compare it against training on each one of the private datasets of the $C$ individual nodes. Moreover, we test whether our unsupervised FL setup can obtain similar accuracy to IID training of the whole dataset (all $280$ speaker IDs available) on a single node.

\subsubsection{Transfer learning with pre-training}
\label{sec:exp_frame:fed_configs:pretrain}
Instead of distributing a random initialization for the global model in the beginning $\w_{g}^{(0)}$, the server could pre-train those weights in order to provide a better initialization point for the FL system (e.g. a big company could provide its clients a pre-trained model using its big data collection to facilitate the federated leraning process). We use the WHAM dataset \cite{Wichern2019WHAM} where we follow a similar training setup as previously explained but now completely supervised and on a single node. After training for $100$ epochs we use those weights as the global model initialization $\w_{g}^{(0)}$ and because we are fine-tuning on the client-side, we lower the learning rate of all local Adam \cite{adam} optimizers to $10^{-4}$. We perform a head-to-head comparison with the same FL setups trained from scratch w.r.t. the benefits that we can get in terms of faster convergence (less communication rounds and data processed on the client-side) as well as the overall performance on the test set.

\subsubsection{Combining supervised and unsupervised clients}
\label{sec:exp_frame:fed_configs:sup_unsup_sweep}
In this experiment we fix the number of clients to $C=64$ and focus on the capability of the FL system to harness the benefits of supervised data which are distributed across clients in a non-IID fashion. In this sense, we combine both \textit{supervised} and \textit{unsupervised} clients where each one performs local training, as explained in Section \ref{sec:method:local_training}, with their local losses $\L_{\operatorname{sup}}$ and $\L_{\operatorname{unsup}}$, respectively. We sweep the proportion of supervised nodes in $p_s \in \{0, 0.25, 0.5, 0.75, 1\}$.

\section{Results and Discussion}
\label{sec:results}

\subsection{Unsupervised federated speech enhancement}
\label{sec:results:unsup_fedenhance}
In Figure \ref{fig:training_from_scratch}, we depict the speech enhancement performance obtained while training for $1 \thinspace 000$ communication rounds. The convergence becomes slower as we increase the number of nodes since aggregating the updates on the weight space becomes harder \cite{FLkairouz2019advancesAndOpenProblems}. However, \fedenhanceSpace{} even with $C=256$ clients significantly outperforms individual training where the dataset is split across $16$ nodes and the local models overfit. Assuming $C=256$ clients, in each communication round $r$, the server needs to aggregate results from $|\mathcal{A}^{(r)}| = \nicefrac{C}{4}$ available clients which accounts for a total memory of $|\w_{c}^{(r)}| \cdot |\mathcal{A}^{(r)}| \approx 204$ MBs. Increasing the number of active clients per round $|\mathcal{A}^{(r)}|$ or increasing the number of local optimization steps per client $K$ transfers the computational load from the communication side to the local audio processing side. From our empirical validation, those values do not seem to affect significantly the overall convergence in terms of communication rounds for our setups with up to $C=256$ clients.

\begin{figure*}[htb!]
  \centering
  \begin{subfigure}[h]{0.5\linewidth}
      \includegraphics[width=\linewidth]{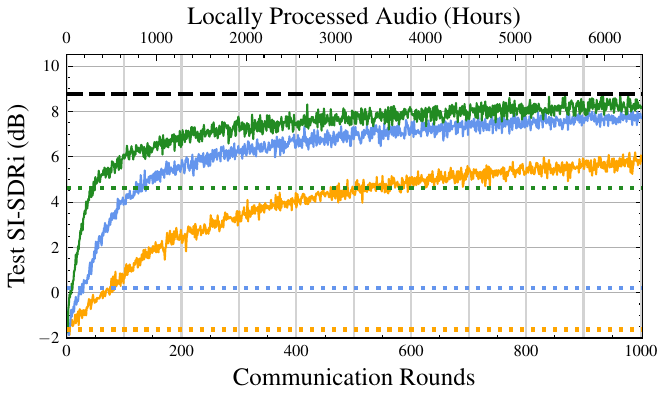}
      \subcaption{Enhancement with one noise source.}
      \label{fig:scratch_1noise} 
     \end{subfigure}
  \begin{subfigure}[h]{0.48\linewidth}
      \includegraphics[width=\linewidth]{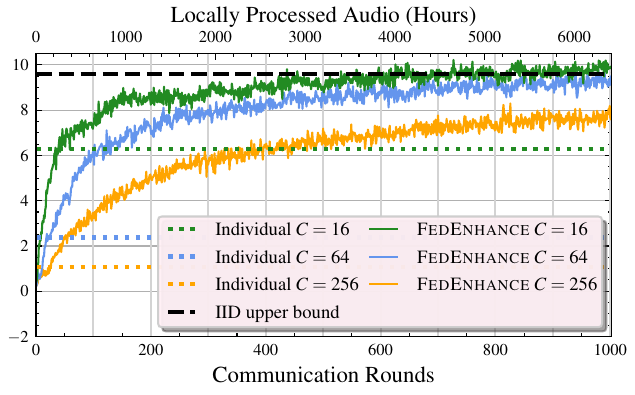}
      \subcaption{Enhancement with two active noise sources.}
      \label{fig:scratch_2noises} 
     \end{subfigure}
    \caption{Speech enhancement performance obtained on \feddataset{} with one noise source (left) and two noise sources (right) versus the total communication rounds between the $C$ clients and the central server as well as the total hours of audio being processed on the client-side. The dotted straight lines symbolize the maximum performance obtained by training in isolation $5$ out of $C$ random clients for $1 \thinspace 000$ epochs on their private data $|\D_c| = \nicefrac{|\D|}{C}$ and taking their mean. The black dashed line on top is the upper bound we could expect assuming that all of the data $\D_c, \enskip c=1, \dots, C$ are available for IID unsupervised training on a single node. Notice that our proposed \fedenhanceSpace{} approach is able to scale to multiple nodes when learning a speech enhancement model using only non-IID and noisy speech recordings.}
\label{fig:training_from_scratch}
\end{figure*} 

\subsection{Transfer learning}
\label{sec:results:pretrained}
By pre-training on the WHAM \cite{Wichern2019WHAM} dataset before distributing the first model from the server to the $C$ clients we see that we can significantly improve the convergence speed of our federated system as shown in Figure \ref{fig:pretraining}. Specifically, using pre-training we close the performance gap between $C=16$ (easy) and $C=256$ (difficult) cases and with less than 100 hours of locally processed audio data. The pre-trained models also achieve higher overall SI-SDRi compared to randomly initialized models on \feddatasetSpace with one or two noise sources present (we omit the Figure for the case with one active noise source since the qualitative results are similar).

\begin{figure}[!htb]
    \centering
      \includegraphics[width=\linewidth]{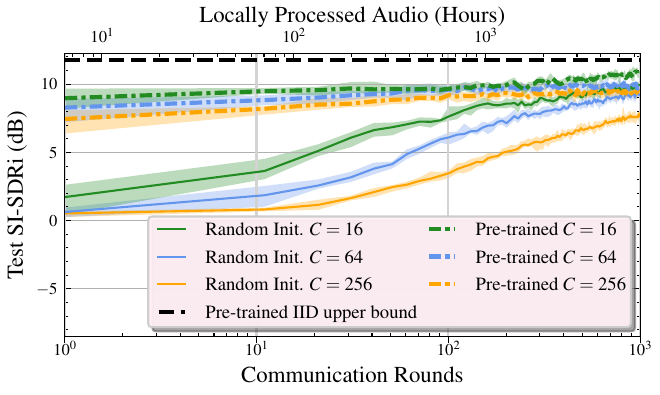}
      \caption{Convergence of \fedenhanceSpace{} on \feddatasetSpace test set with two active noise sources when starting from the pre-trained model (dotted) and when starting from a random initialization (solid). For better visualization, we plot the mean across $10$ communication rounds in log scale while the shaded regions are defined by the minimum and maximum SI-SDRi obtained in the same number of rounds. The black dashed line on top is the upper bound obtained with IID unsupervised training on a single node starting from the same pre-trained model.}
      \label{fig:pretraining}
\end{figure}

\subsection{Combining supervised and unsupervised clients}
\label{sec:results:sup_unsup}
In Figure \ref{fig:psup_sweep}, we show the performance obtained with \fedenhanceSpace{} with $C=64$ clients for \feddatasetSpace{} validation and test sets and with the presence of one or two noise sources after $1 \thinspace 000$ rounds. We select the model with the highest SI-SDRi on the validation set over the last $50$ communication rounds. Notice that in the case of all clients having access to supervised data (rightmost) we can obtain a Test SI-SDRi of $9.3$dB and $10.9$dB for one and two noise sources, respectively. By assuming $50\%$ of supervised clients, \fedenhance{} can leverage the supervised information flow and improve upon the totally unsupervised configuration (leftmost) in terms of Test SI-SDRi ($8.0 \rightarrow 8.9$dB and $9.4 \rightarrow 10.5$dB for one and two noise sources, respectively).

\begin{figure}[!htb]
    \centering
      \includegraphics[width=\linewidth]{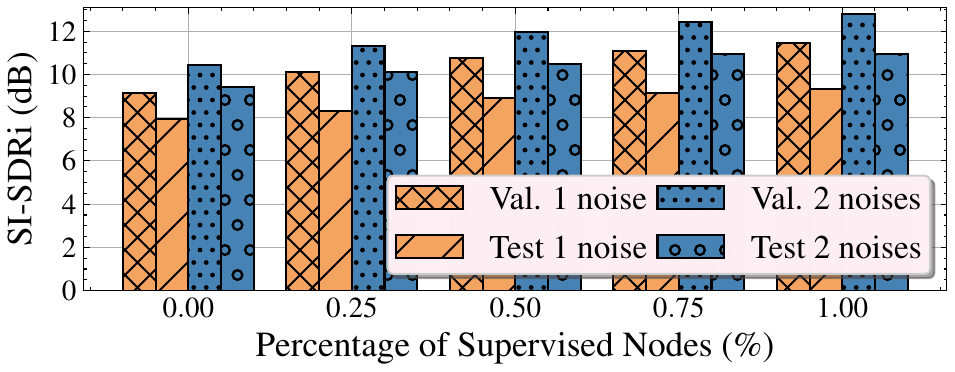}
      \caption{Speech enhancement performance on \feddatasetSpace validation and test sets with one or two active noise sources while sweeping the number of supervised nodes from totally unsupervised FL (left) to totally supervised FL (right).}
      \label{fig:psup_sweep}
\end{figure}

\section{Conclusion}
\label{sec:conclusion}
We have presented an unsupervised federated learning approach, namely, \fedenhance, which is capable of collectively training a speech enhancement model with non-IID distributed noisy speech recordings across multiple clients. Our experiments show that our approach is able to obtain competitive performance with IID training on a single node and can be further boosted by using pre-training on other datasets as well as by combining nodes with supervised data. In the future, we aim to extend our approach to multi-task problems including simultaneous sound recognition and separation.

\newpage
\bibliographystyle{IEEEtran}
\bibliography{refs21}

\end{sloppy}
\end{document}